\documentclass[11pt,submission]{IEEEtran}
\newtheorem{theorem}{Theorem}

\title{General Classes of Lower  Bounds on the Probability of Error
in Multiple Hypothesis Testing}
\author{Tirza Routtenberg and Joseph Tabrikian \\ Department of Electrical and Computer Engineering,  \\ Ben-Gurion University of the Negev
 Beer-Sheva 84105, Israel \\
 Phone: $+$\,972\,-86477774, Email: \{tirzar,joseph\}@ee.bgu.ac.il}

\usepackage{amssymb, amsmath}
\usepackage{graphicx,cite}
 \usepackage[dvips]{color}
 \newcommand{\ud}{\,\mathrm{d}}

\newcommand{\xvec}{{\bf{x}}}

\newcommand{\zetavec}{{\bf{\zeta}}}

\newcommand{\define}{\stackrel{\triangle}{=}}

\newcommand{\be}{\begin{equation}}
\newcommand{\ee}{\end{equation}}
\newcommand{\beqna}{\begin{eqnarray}}
\newcommand{\eeqna}{\end{eqnarray}}

\input{psfig.sty}
\begin{document}
\maketitle
\nopagebreak

\begin{abstract}
In this paper, two new classes of lower bounds on the probability of error for $m$-ary hypothesis testing are
proposed.
Computation of the
minimum probability of error which is attained  by the maximum
{\em a-posteriori} probability (MAP) criterion,
is usually not tractable.
The new classes are derived using H$\ddot{\text{o}}$lder's inequality and reverse H$\ddot{\text{o}}$lder's inequality. The bounds in these classes  provide good prediction of
the minimum probability of error in multiple hypothesis testing.  The new classes generalize and extend existing bounds and their relation  to some existing upper bounds is presented.
It is shown that    the tightest bounds in these classes  asymptotically coincide
with the optimum probability of error provided by the MAP criterion for binary or multiple hypothesis testing problem. These bounds are compared
with other existing lower bounds in several typical detection and classification problems in terms
of tightness and computational complexity.
\end{abstract}

\section{Introduction}
\label{sec:intro}

 Lower bounds on the probability of error are of great importance in system design and
performance analysis in many applications, such as signal detection, communications \cite{Gallager}, 
classification, and pattern recognition \cite{PRbook}. It is well known that the minimum probability of error is attained by the maximum
{\em a-posteriori} probability (MAP) criterion, however, its probability of error is often difficult to
calculate and usually not tractable. In such cases, lower bounds on the probability of error are
useful for performance analysis, feasibility study and system design. These bounds can be useful
also for derivation of analytical expressions for the  family of Ziv-Zakai lower  bounds (ZZLB) for parameter
estimation \cite{ZZLB}. One of the difficulties in computation of the ZZLB is that they involve
an expression for the minimum probability of error of a binary hypothesis problem. Analytic expressions for
lower bounds on the probability of error may be useful to simplify the calculation of the bound.
Another application of 
these bounds  is a sphere packing lower bound on
probability of error under MAP of the ensemble of random
codes  \cite{Renyis}.

Several lower bounds on the probability of error have been presented in the literature,
 for specific problems, such as  signals in white Gaussian noise \cite{WGN}, \cite{WGN2}, and for general statistical models.
 The general bounds can be divided into bounds for  binary hypothesis problems
{\cite{tutorial,Boekee86, Chernoff, Avi96, Hashlamoun_doc}} and   bounds for multiple-hypothesis problems  \cite{Renyis}, {\cite{Shannon, fano, feder, HR70,LBpoor,  pattern, HVLB, Matusita, boekee79, Basseville89}}.
Several lower and upper bounds  utilize
distance measures between statistical distributions, like
 Bhattacharyya distance   \cite{tutorial}, \cite{Boekee86},  Chernoff \cite{Chernoff},  Bayesian distance \cite{pattern}, Matusita distance \cite{Matusita}, and the general mean distance \cite{boekee79}, \cite{Basseville89}.
Two classical lower bounds on the multiple-hypothesis error probability
that have been used in proving coding theorems are the    Shannon  \cite{Shannon}  and Fano \cite{fano} inequalities. The relations between entropy and error probability have been used to derive the bounds in
 {\cite{feder,Renyis,HR70}}. The bound in  \cite{LBpoor} has been derived using  proofs of  converse channel
coding theorems in information theory.
 In addition, there are several ad-hoc binary hypothesis testing bounds that directly bound  the minimum function on the {\em a-posteriori} probabilities. This class includes the ``Gaussian-Sinusoidal'' upper and lower bounds \cite{Hashlamoun_doc} and the exponential bound \cite{Avi96}, that  are found to be useful in some specific cases.   A brief
review of some existing lower bounds on the probability of error is presented in Appendix B and in \cite{error_eilat}.

Practical and useful lower bounds on the probability of error are expected to
be computationally simple, tight, and appropriate for general multi-hypothesis problems.
In this paper, two new classes of lower bounds with the aforementioned desired properties are
derived using H$\ddot{\text{o}}$lder's inequality and reverse H$\ddot{\text{o}}$lder's inequality. The bounds in these classes  provide good prediction of
the minimum probability of error in multiple hypothesis testing and are often
easier to evaluate than the MAP probability of error. These bounds are compared
with other existing lower bounds.  In addition, it is shown that the new classes generalize some existing lower bounds \cite{Renyis}, \cite{pattern}, {\cite{Basseville89, TOUdoc, Vaj68}}. 
 The tightest lower bound under each
class of bounds is derived and it is shown that the tightest bound  asymptotically coincides
with the optimum probability of error provided by the MAP criterion. 

The paper is organized as follows. 
The new classes of bounds are derived in Section
\ref{the_bound} and the bounds properties are presented in Section \ref{prop}. In Section \ref{App}, simple versions of the ZZLB for parameter estimation are derived using the proposed classes of bounds.
The  performances of the proposed bounds for
various examples is evaluated in Section \ref{simulations}. Finally, our conclusions appear in Section
 \ref{diss}.

\section{General  classes of bounds on probability   of error}
\label{the_bound}
 \subsection{Problem statement}
 \label{problem}
Consider an $M$-ary hypothesis testing problem, in which the hypotheses are $\theta_i,~~~i=1,\ldots,M$ with the corresponding 
 {\em a-priori} probabilities $P(\theta_i),~~~i=1,\ldots,M$.
 Let $P(\theta_i|\xvec),~~~i=1,\ldots,M$ denote the conditional probability of $\theta_i$ given the random observation vector,  $\xvec$. 
The probability of error of the decision problem is denoted by $P_e$.
It is well known that the minimum  probability of error obtained by the MAP criterion,
is given by \cite{feder}
\begin{equation}
\label{Pe_MAP}
P_e^{(min)}=1-{\rm E}\left[\max\sb{i=1,\ldots,M}P(\theta_i|\xvec)\right] 
\end{equation}
where the MAP detector is \[\hat{\theta}_{MAP}=\arg\max\sb{\theta\in \{\theta_1,\ldots,\theta_M\}}P(\theta|\xvec) \;.\]
However, the minimum probability of error in (\ref{Pe_MAP}) is often difficult to
calculate and usually not tractable. Therefore, 
computable and tight lower and upper bounds on the probability of error  are useful for performance
analysis and system design.
\subsection{Derivation of the general classes of bounds}
Consider the above  $M$-ary hypothesis testing problem with detector $\hat{\theta}=\hat{\theta}(\xvec)$. The detector $\hat{\theta}(\xvec)$ is assumed to be an arbitrary detector that decides on one of the hypotheses with positive (non-zero)  {\em a-posteriori} probabilities. That is, in the case where  $P(\theta_j|\xvec)=0$ we assume that $\hat{\theta}(\xvec) \neq \theta_j$ with probability 1 (w.p.1). Let
\begin{equation}
\label{Udef}
u(\xvec,\theta)\define\mathbf{1}_{\hat{\theta}\neq\theta}= \left\{
\begin{array}{rl}
1 & \text{if }\hat{\theta}\neq\theta\\
0 & \text{if } \hat{\theta}=\theta
\end{array} \right. ,
\end{equation}
where $\theta$ is the true hypothesis. It can be verified  that
\begin{equation}
\label{pe_def}
P_e={\rm E}\left[ u(\xvec,\theta)\right]={\rm E}\left[ |u(\xvec,\theta)|^p\right]
\end{equation}
and
\begin{equation}
\label{pe_def2}
P_e =1-{\rm E}\left[ 1-u(\xvec,\theta)\right]=1-{\rm E}\left[ |1-u(\xvec,\theta)|^{p}\right]
\end{equation}
for every $p>0$, where $P_e$ is the probability of error of the detector $\hat{\theta}$.
Then, according to  H$\ddot{\text{o}}$lder's inequality and reverse H$\ddot{\text{o}}$lder's inequality \cite{error_eilat},  \cite{HLP}:
\begin{equation}
\label{ineq}
{\rm E}^{\frac{1}{p}}\left[ |u(\xvec,\theta)|^p\right]{\rm E}^{\frac{p-1}{p}}\left[ |v_1(\xvec,\theta)|^{\frac{p}{p-1}}\right]\geq {\rm E}\left[ |u(\xvec,\theta)v_1(\xvec,\theta)|\right],~~~ p>1
\end{equation}
and

\begin{equation}
\label{rev_ineq}
 {\rm E}\left[ |\left(1-u(\xvec,\theta)\right)v_2(\xvec,\theta)|\right]\geq {\rm E}^{p}\left[ |1-u(\xvec,\theta)|^{\frac{1}{p}}\right]{\rm E}^{1-p}\left[ |v_2(\xvec,\theta)|^{\frac{1}{1-p}}\right],~~~p>1
\end{equation}
for  arbitrary scalar functions $v_1(\xvec,\theta)$ and $v_2(\xvec,\theta)$.

By substituting of  (\ref{pe_def}) and (\ref{pe_def2}) into
(\ref{ineq})  and
(\ref{rev_ineq}), respectively, one obtains the following lower bounds on the  probability of error:
\begin{equation}
\label{LB}
P_e\geq {\rm E}^p\left[ |u(\xvec,\theta)v_1(\xvec,\theta)|\right]{\rm E}^{1-p}\left[ |v_1(\xvec,\theta)|^{\frac{p}{p-1}}\right],~~~ p> 1
\end{equation}
\begin{equation}
\label{LB2}
 P_e\geq 1- {{\rm E}^{\frac{1}{p}}\left[ |\left(1-u(\xvec,\theta)\right)v_2(\xvec,\theta)|\right]}{{\rm E}^{\frac{p-1}{p}}\left[ |v_2(\xvec,\theta)|^{{\frac{1}{1-p}}}\right]},~~~p>1 \;.
\end{equation}
By substituting different functions $v_1(\xvec,\theta),~v_2(\xvec,\theta)$ in (\ref{LB})-(\ref{LB2}), one obtains different lower bounds on the probability of error. In general, this bound is a function of  the   detector via $u(\xvec,\theta)$.  
The following theorem states the condition to obtain valid bounds which are independent of the estimator.
\begin{theorem}
\label{Th1}
Under the assumption that $P(\theta_i|\xvec)>0~~~\forall \xvec\in\chi$ and $\theta_i,~ i=1,...,M$,
a necessary and sufficient condition   to obtain a valid bound on the   probability of error which is independent of the detector $\hat{\theta}$, is that the functions $v_1\left(\xvec,\theta\right)$ and $v_2\left(\xvec,\theta\right)$ should be structured as follows
\begin{equation}
\label{Vdef}
 v_k(\xvec,\theta_i)=\frac{{\zetavec_k}(\xvec)}{P(\theta_i|\xvec)}, ~~~k=1,2,~~i=1,\ldots,M
\end{equation}
where $\zetavec_1(\cdot)$ and $\zetavec_2(\cdot)$ are  arbitrary functions of the observations $\xvec$ and
with no loss of generality they should be chosen to be  non-negative.
  \end{theorem}
  {\em{Proof:}} In  Appendix A.

Using (\ref{Vdef}) it is shown in Appendix A that using $\{v_k(\xvec,\theta_i)\}_{k=1,2}$ defined in (\ref{Vdef})
\begin{eqnarray}
\label{num2}
{\rm E}\left[ |u(\xvec,\theta) v_1\left(\xvec,\theta\right)|\right]
=(M-1){\rm E}\left[\zetavec_1(\xvec)  \right],
\end{eqnarray}
\begin{eqnarray}
\label{num3}
{\rm E}\left[ |\left(1-u(\xvec,\theta) \right)v_2\left(\xvec,\theta\right)|\right]
={\rm E}\left[\zetavec_2(\xvec)  \right],
\end{eqnarray}
and
\begin{eqnarray}
\label{den}
{\rm E}\left[ \left|v_1\left(\xvec,\theta\right)\right|^{\frac{p}{p-1}}\right]
={\rm E}\left[{\zetavec}_1^{\frac{p}{p-1}}(\xvec)  \sum\limits_{i=1}^M{P^{{\frac{1}{1-p}}}\left(\theta_i|\xvec\right)} \right],
\end{eqnarray}
\begin{eqnarray}
\label{den2}
{\rm E}\left[ \left|v_2\left(\xvec,\theta\right)\right|^{\frac{1}{1-p}}\right]
={\rm E}\left[{\zetavec}_2^{\frac{1}{1-p}}(\xvec)  \sum\limits_{i=1}^M{P^{{\frac{p}{p-1}}}\left(\theta_i|\xvec\right)} \right]\;.
\end{eqnarray}
By substitution of (\ref{num2}), (\ref{den})  into (\ref{LB}), and (\ref{num3}), (\ref{den2})  into (\ref{LB2}),   the  new classes of lower bounds    can be rewritten as:
\begin{equation}
\label{lowerBB}
P_e \geq (M-1)^p {\rm E}^p\left[\zetavec_1(\xvec)  \right]
{\rm E}^{1-p}\left[\zetavec_1^{\frac{p}{p-1}}(\xvec)\sum\limits_{i=1}^M{P^{\frac{1}{1-p}}\left(\theta_i|\xvec\right)}  \right],~~~ p>1
\end{equation}
\begin{equation}
\label{lowerBBB}
P_e\geq 1- {{\rm E}^{\frac{1}{p}}\left[\zetavec_2(\xvec) \right]}{\rm E}^{\frac{p-1}{p}}\left[{\zetavec}_2^{\frac{1}{1-p}}(\xvec)  \sum\limits_{i=1}^M{P^{{\frac{p}{p-1}}}\left(\theta_i|\xvec\right)} \right],~~~p>1 \;.
\end{equation}

\subsection{The tightest subclasses in the proposed classes of lower bounds}
According to   H$\ddot{\text{o}}$lder's inequality \cite{HLP}
\beqna
{\rm E}^p\left[\zetavec_1(\xvec)  \right]
{\rm E}^{1-p}\left[\zetavec_1^{\frac{p}{p-1}}(\xvec)\sum\limits_{i=1}^M{P^{{\frac{1}{1-p}}}\left(\theta_i|\xvec\right)}  \right]&\geq&
{\rm E}\left[\zetavec_1^p(\xvec) 
\left(\zetavec_1^{\frac{p}{p-1}}(\xvec)\sum\limits_{i=1}^M{P^{\frac{1}{1-p}}\left(\theta_i|\xvec\right)}  \right)^{1-p}\right]\nonumber\\&=&{\rm E}\left[ \left(\sum\limits_{i=1}^M{P^{\frac{1}{1-p}}\left(\theta_i|\xvec\right)} \right)^{1-p} \right]
\eeqna
and it becomes an equality {\em{iff}}
\be
\label{zeta1}
\zetavec_1(\xvec)=c_1\zetavec_1^{{\frac{p}{p-1}}}(\xvec)\sum\limits_{i=1}^M{P^{\frac{1}{1-p}}\left(\theta_i|\xvec\right)} 
\ee
where $c_1$ denotes a constant independent of $\xvec$ and $\theta_i,~i=1,\ldots,M$. 
In similar,
\beqna
{{\rm E}^{\frac{1}{p}}\left[\zetavec_2(\xvec) \right]}{\rm E}^{\frac{p-1}{p}}\left[{\zetavec}_2^{\frac{1}{1-p}}(\xvec)  \sum\limits_{i=1}^M{P^{{\frac{p}{p-1}}}\left(\theta_i|\xvec\right)} \right]&\geq& {\rm E}\left[\zetavec_2^{p}(\xvec) \left(\zetavec_2^{{\frac{1}{1-p}}}(\xvec)\sum\limits_{i=1}^M{P^{\frac{p}{p-1}}\left(\theta_i|\xvec\right)} \right)^{\frac{p-1}{p}} \right]\nonumber\\&=&{\rm E}\left[ \left(\sum\limits_{i=1}^M{P^{\frac{p}{p-1}}\left(\theta_i|\xvec\right)} \right)^{\frac{p-1}{p}} \right]
\eeqna
and it becomes an equality {\em{iff}}
\be
\label{zeta2}
\zetavec_2(\xvec)=c_2{\zetavec}_2^{\frac{1}{1-p}}(\xvec)  \sum\limits_{i=1}^M{P^{{\frac{p}{p-1}}}\left(\theta_i|\xvec\right)} 
\ee
where $c_2$ denotes a constant independent of $\xvec$ and $\theta_i,~i=1,\ldots,M$. 
Thus, the tightest subclasses of bounds in the two classes are:
\begin{eqnarray}
\label{final_bound}
P_e\geq B_p^{(1)}\define
(M-1)^{p}{\rm E}\left[\left(\sum\limits_{i=1}^M P^{\frac{1}{1-p}}\left(\theta_i|\xvec\right)\right)^{1-p}\right],~~~\forall p>1
\end{eqnarray}
\begin{eqnarray}
\label{final_bound2}
P_e \geq B_p^{(2)}\define 1-{\rm E}\left[ \left(\sum\limits_{i=1}^M{P^{\frac{p}{p-1}}\left(\theta_i|\xvec\right)} \right)^{\frac{p-1}{p}} \right],~~~\forall p>1
\end{eqnarray} 
obtained by substituting (\ref{zeta1}) and (\ref{zeta2})
in (\ref{lowerBB}) and (\ref{lowerBBB}), respectively. 

\subsection{Simplifications of the bound}
\label{J_ineq}
The bounds in (\ref{final_bound}) and (\ref{final_bound2}) can be simplified  using Jensen's inequality \cite{Rudin}. 
Let $\varepsilon_1=\sum\limits_{i=1}^M P^{\frac{1}{1-p}}\left(\theta_i|\xvec\right)>0$, than for $p>1$
 $ B_p^{(1)}=\varepsilon_1^{1-p}$  is a convex function
of  $\varepsilon_1>0$.
According to Jensen's inequality for convex functions 
\begin{eqnarray}
\label{Jfinal_bound}
B_p^{(1)}=
(M-1)^{p}{\rm E}\left[\left(\sum\limits_{i=1}^M P^{\frac{1}{1-p}}\left(\theta_i|\xvec\right)\right)^{1-p}\right]\geq JB_p^{(1)}\define
(M-1)^{p}{\rm E}^{1-p}\left[\sum\limits_{i=1}^M P^{\frac{1}{1-p}}\left(\theta_i|\xvec\right)\right],~~~\forall p>1\;.
\end{eqnarray}

In similar,  $B_p^{(2)}=\varepsilon_2^{\frac{p-1}{p}}$ is a concave function of the positive term $\varepsilon_2=\sum\limits_{i=1}^M{P^{\frac{p}{p-1}}\left(\theta_i|\xvec\right)}$.
According to Jensen's inequality for concave functions
\begin{eqnarray}
\label{Jfinal_bound2}
B_p^{(2)}= 1-{\rm E}\left[ \left(\sum\limits_{i=1}^M{P^{\frac{p}{p-1}}\left(\theta_i|\xvec\right)} \right)^{\frac{p-1}{p}} \right]
\geq JB_p^{(2)}\define 1-{\rm E}^{\frac{p-1}{p}}\left[ \sum\limits_{i=1}^M{P^{\frac{p}{p-1}}\left(\theta_i|\xvec\right)} \right],~~~\forall p>1\;.
\end{eqnarray} 

\section{Properties of the proposed classes of  bounds}
\label{prop}
\subsection{Asymptotic properties}
According to \cite{HLP} (Theorem 19, page 28), for any sequence of nonnegative numbers, $a_1,\ldots,a_M$
\be
\label{HLP2}
\left(\sum_{i=1}^M a_i^s\right)^{\frac{1}{s}}\geq \left(\sum_{i=1}^M a_i^t\right)^{\frac{1}{t}},~~~\forall 0<s<t 
\ee
 and thus,
the term within the expectation in (\ref{final_bound}) \[\left(\sum\limits_{i=1}^M P^{\frac{1}{1-p}}\left(\theta_i|\xvec\right)\right)^{1-p}=\frac{1}{\left(\sum\limits_{i=1}^M \left(\frac{1}{P\left(\theta_i|\xvec\right)}\right)^{\frac{1}{p-1}}\right)^{p-1}}\] 
is a decreasing function of $p$ for all $p>1$. 
Therefore, in the binary case,  the bound in (\ref{final_bound}) satisfies
\be
\label{norm_ineq}
B_p^{(1)}\geq B_r^{(1)},~~~\forall 1<p\leq r,~~~M=2\;.
\ee
In particular, for $p\rightarrow 1^+$, the bound in (\ref{final_bound}) becomes
\begin{eqnarray}
B_{1^+}^{(1)}=\lim_{p\rightarrow 1^+}B_p^{(1)}={\rm E}\left[\min\sb{i=1,2}  P\left(\theta_i|\xvec\right)\right]=1-{\rm E}\left[\max\limits\sb{i=1,2}  P\left(\theta_i|\xvec\right)\right],
\end{eqnarray}
which is the tightest  lower bound on the probability of error  in the first proposed class of lower bounds for $M=2$.
 Thus, for the binary hypothesis testing the  bound  in (\ref{final_bound}) with $p\rightarrow1^+$ is tight and
attains the minimum probability of error, presented in (\ref{Pe_MAP}).

In similar, using (\ref{HLP2})
 the term  $\left(\sum\limits_{i=1}^M{P^{{\frac{p}{p-1}}}\left(\theta_i|\xvec\right)}\right)^{\frac{p-1}{p}}$ from (\ref{final_bound2}), which is the $\frac{p}{p-1}$ norm of $\left\{P( \theta_i|\xvec) \right\}_{i=1,\ldots M}$,
is a decreasing function of $p$ for all $p>1$. 
 Therefore, in the general case
\be
\label{norm_ineq2}
B_p^{(2)}\geq B_r^{(2)},~~~\forall 1<p\leq r,~~~\forall M\;.
\ee
In particular, for $p\rightarrow 1^+$, the bound in (\ref{final_bound2}) becomes
\begin{eqnarray}
B_\infty^{(2)}=\lim_{p\rightarrow 1^+}B_p^{(2)}= 1-{\rm E}\left[\max\sb{i=1,\ldots,M} P\left(\theta_i|\xvec\right)\right]
\end{eqnarray}
which is the minimum  probability of error, obtained by the MAP criterion.
 Thus, for the M-hypothesis testing, the  bound  in (\ref{final_bound2}) with $p\rightarrow 1^+$ is tight and
attains the minimum probability of error presented in (\ref{Pe_MAP}).

\subsection{Generalization of existing bounds}
In this section, we show that the proposed classes of bounds generalize some existing bounds. 
In particular, the  lower bounds in \cite{Renyis}, \cite{pattern}, and \cite{Basseville89} can be interpreted as  special cases of the proposed general $M$-hypotheses bounds, presented in (\ref{final_bound}) and (\ref{final_bound2}).
In the binary hypothesis testing,  the bound in (\ref{final_bound}) with $p=2$ can be written by the following simple version:
\begin{eqnarray}
\label{LBq2}
P_e\geq
{\rm E}\left[P\left(\theta_1|\xvec\right)P\left(\theta_2|\xvec\right)\right]
\end{eqnarray}
which is identical to
the harmonic lower bound \cite{Renyis} and to the Vajda's
quadratic entropy bound \cite{TOUdoc}, \cite{Vaj68} with $M=2$.

In the multiple hypothesis testing  the bound in (\ref{final_bound2}) with $p=2$ can be written in the following simple version:
\begin{eqnarray}
\label{LB2q2}
P_e\geq
1-{\rm E}\left[\sqrt{\sum_{i=1}^{M}P^2(\theta_i|\xvec)}\right]
\end{eqnarray}
which is identical to the Bayesian lower bound \cite{Renyis}, \cite{pattern}, $B^{(Bayes3)}$,    described in Appendix B. The  bound $JB_p^{(2)}$ in (\ref{Jfinal_bound2}) with $p=2,~M=2$ is
\begin{eqnarray}
P_e\geq
JB_2^{(2)}=1-\sqrt{{\rm E}\left[\sum_{i=1}^{M}P^2(\theta_i|\xvec)\right]}
\end{eqnarray}
is identical to the Bayesian lower bound \cite{pattern}, $B^{(Bayes2)}$,  described in Appendix B.
In addition,
in the multiple hypothesis testing, the subclass of bounds in (\ref{final_bound2}) is a   ``general mean distance" class of bounds presented   in Appendix B in (\ref{Gab2}).

\subsection{Relation to upper bounds on minimum probability of error}
In \cite{Renyis}, a class of  upper bounds on the MAP probability of error for binary hypothesis testing is derived using the negative power mean inequalities:
\begin{equation}
\label{upper bound}
P_e^{(min)} \leq 2^{(p-1)}{\rm E}_{\xvec}\left[\left(\sum\limits_{i=1}^2  P\left(\theta_i|\xvec\right)^{\frac{1}{1-p}}\right)^{1-p}\right]
\end{equation}
for any $p>1$. It can be seen that this class of upper bounds is proportional to the proposed tightest subclass of lower bound in (\ref{final_bound}) with a factor of $2^{p-1}$. This factor  controls the
tightness between upper and lower bounds in the probability
of error for binary hypothesis testing. This upper bound coincides with the proposed lower bound $B_p^{(1)}$ in the limit of $p\rightarrow 1^+$.

In \cite{Basseville89}, the ``general mean distance" is used in order to derive upper bounds on the MAP probability of error. One subclass of upper bounds presented in this reference is
\beqna
\label{Gab3}
P_e^{(min)}\leq B_p^{(GMD3)}=1-M^{\frac{1-p}{p}}{\rm E}\left[ \left(\sum_{i=1}^M P^{\frac{p}{p-1}}\left(\theta_i|\xvec\right)\right)^{\frac{p-1}{p}}\right],~~~p>1 \;.
\eeqna
It can be seen that $1-B_p^{(GMD3)}=M^{\frac{1-p}{p}}\left(1-B_p^{(2)}\right)$. Thus, 
the computations of  (\ref{Gab3}) for specific hypothesis testing problems can be utilized to compute the lower bounds in (\ref{final_bound}) for the same problems.

%%%%%%%%%%%%%%%%%%%%%%%%%%%%%%%%%%%%%%%%%%%%%%%%%%%%%

%%%%%%%%%%%%%%%%%%%%%%%%%%%%%%%%%%%%%%%%%%%%%%%%%%%%%%%%%%%%%%%%%%%%%%%%%%%%%%%%%%%%%%%%%%%%%%%%%%%%%%%%%%%%%%%
\section{Application:  simple versions of the Ziv-Zakai lower bound}
\label{App}
The new classes of lower bounds on the probability of error
can be  used to
derive simple closed  forms of the ZZLB for  Bayesian parameter estimation. A critical factor in implementing the extended ZZLB
is the evaluation of the probability of error in  a binary detection problem.
The bounds are useful only if the probability of error is known or can be
tightly lower bounded. Thus, the new classes of lower bounds in (\ref{lowerBB}), (\ref{lowerBBB}) can be used in order to derive lower bounds on the ZZLB, providing less tighter MSE bounds which may be easier to compute. Note that the derivation in this section is performed under the assumption that $\xvec$ is continuous random variable. Extension to any random variable $\xvec$  with ${\rm{E}}[\xvec^2]<\infty$ is straightforward.

Consider the estimation of a continuous scalar random variable $\phi\in\Phi$, with {\em a-priori} probability density function (pdf) $f_\phi(\phi)$,
based on an observation vector $\xvec\in\chi$. The pdf's $f_{\phi|\xvec}(\cdot|\xvec)$  denotes the
conditional pdf of $\phi$ given $\xvec$. For any estimator $\hat{\phi}(\xvec)$ with estimation error  $\epsilon=\hat{\phi}(\xvec)-\phi$, the mean-square-error (MSE)  is defined as
${\rm E}\left[\left|\hat{\phi}(\xvec)-\phi\right|^2\right]$.
The extended ZZLB  is \cite{ExtendedZZ}
\begin{eqnarray}
\label{ZZLB_A}
{\rm E}\left[\left|\hat{\phi}(\xvec)-\phi\right|^2\right]\geq ZZLB= \frac{1}{2}\int\limits_0^\infty V\left\{\int\limits_{-\infty}^{\infty}\left(f_\phi(\varphi)+f_\phi(\varphi+h)\right) P_{min}(\varphi,\varphi+h){\ud} \varphi\right\}h {\ud} h
\end{eqnarray}
where $P_{min}(\varphi,\varphi+h)$ is the minimum probability of error for the following detection problem:
\begin{equation}
\label{detection_prob}
{\begin{array}{rl}
&H_0: f_{\xvec|H_0}(\xvec)= f_{\xvec|\phi}(\xvec|\varphi)\\
&H_1: f_{\xvec|H_1}(\xvec)= f_{\xvec|\phi}(\xvec|\varphi+h)
\end{array}}
\end{equation}
 with prior probabilities
\be
\label{priorP}
P(H_0)=\frac{f_{\phi}(\varphi)}{f_{\phi}(\varphi)+f_{\phi}(\varphi+h)},~~~P(H_1)=1-P(H_0)\;.
\ee
The operator $V$ returns
a nonincreasing function  by filling
in any valleys in the input function
\begin{equation}
Vf(h)=\max\sb{\xi\geq 0}f(h+\xi),~~~h\in{\mathbb{R}}\;.
\end{equation}

Since $f_\phi(\varphi),~f_\phi(\varphi+h)$, and $P_{min}(\varphi,\varphi+h)$ are non-negative terms, the inner integral term in  (\ref{ZZLB_A}) can be lower bounded by bounding $ P_{min}(\varphi,\varphi+h)$.
Thus, 
\begin{eqnarray}
\label{ZZLB_B}
{\rm E}\left[\left|\hat{\phi}(\xvec)-\phi\right|^2\right]\geq ZZLB\geq C_p\define \frac{1}{2}\int\limits_0^\infty V\left\{\int\limits_{-\infty}^{\infty}\left(f_\phi(\varphi)+f_\phi(\varphi+h)\right) LB(\varphi,h){\ud} \varphi\right\}h {\ud} h
\end{eqnarray}
where $LB(\varphi,h)$ is any lower bound on the minimum probability of error of the detection problem stated in (\ref{detection_prob}).  By substituting  the lower bound on the probability of outage error from  (\ref{lowerBB}) and (\ref{lowerBBB}), respectively, in (\ref{ZZLB_B}) with $M=2$ and using arbitrary non-negative functions  $\zetavec_1(\xvec)$, $\zetavec_2(\xvec)$ 
one obtains different MSE lower bounds.
In particular, by substituting $LB=B_p^{(1)}$ and $LB=B_p^{(2)}$ from 
 (\ref{final_bound}) and (\ref{final_bound2}), respectively in (\ref{ZZLB_B}),
one obtains the tightest classes of MSE lower bounds 
\begin{eqnarray}
\label{ZZLB_C}
{\rm E}\left[\left|\hat{\phi}(\xvec)-\phi\right|^2\right]\geq C_p^{(1)}\define \frac{1}{2}\int\limits_0^\infty V\left\{\int\limits_{-\infty}^{\infty} {\rm E}\left[ \left(f_{\phi|\xvec}^{\frac{1}{1-p}}\left(\varphi|\xvec\right)+f_{\phi|\xvec}^{\frac{1}{1-p}}\left(\varphi+h|\xvec\right)\right)^{1-p} \right]{\ud} \varphi\right\}h {\ud} h ,~~~\forall p>1
\end{eqnarray}
and
\begin{eqnarray}
\label{ZZLB_D}
{\rm E}\left[\left|\hat{\phi}(\xvec)-\phi\right|^2\right]\geq C_p^{(2)}\define
\frac{1}{2}\int\limits_0^\infty V\left\{  2-\int\limits_{-\infty}^{\infty} {\rm E}\left[ \left(f_{\phi|\xvec}^{\frac{p}{p-1}}\left(\varphi|\xvec\right)+f_{\phi|\xvec}^{\frac{p}{p-1}}\left(\varphi+h|\xvec\right)\right)^{\frac{p-1}{p}} \right]{\ud}  \varphi\right\}h {\ud} h\nonumber\\=
\frac{1}{2}\int\limits_0^\infty V\left\{  \int\limits_{-\infty}^{\infty} {\rm E}\left[f_{\phi|\xvec}\left(\varphi|\xvec\right)+f_{\phi|\xvec}\left(\varphi+h|\xvec\right)- \left(f_{\phi|\xvec}^{\frac{p}{p-1}}\left(\varphi|\xvec\right)+f_{\phi|\xvec}^{\frac{p}{p-1}}\left(\varphi+h|\xvec\right)\right)^{\frac{p-1}{p}} \right]{\ud}  \varphi\right\}h {\ud} h,
\end{eqnarray}
$\forall p>1$.
For $p\rightarrow 1^+$, the bounds in (\ref{ZZLB_C}) and (\ref{ZZLB_D}) become
\begin{eqnarray}
{\rm E}\left[\left|\hat{\phi}(\xvec)-\phi\right|^2\right]\geq
\frac{1}{2}\int\limits_0^\infty V\left\{  \int\limits_{-\infty}^{\infty} {\rm E}\left[\min \left(f_{\phi|\xvec}\left(\varphi|\xvec\right),f_{\phi|\xvec}\left(\varphi+h|\xvec\right)\right) \right]{\ud}  \varphi\right\}h {\ud} h,
\end{eqnarray}
which coincides with the ZZLB as presented in \cite{Belldoc}.

\section{Examples}
\label{simulations}
\subsection{Bounds comparison}
\begin{figure}[htb]
\centerline{\psfig{figure=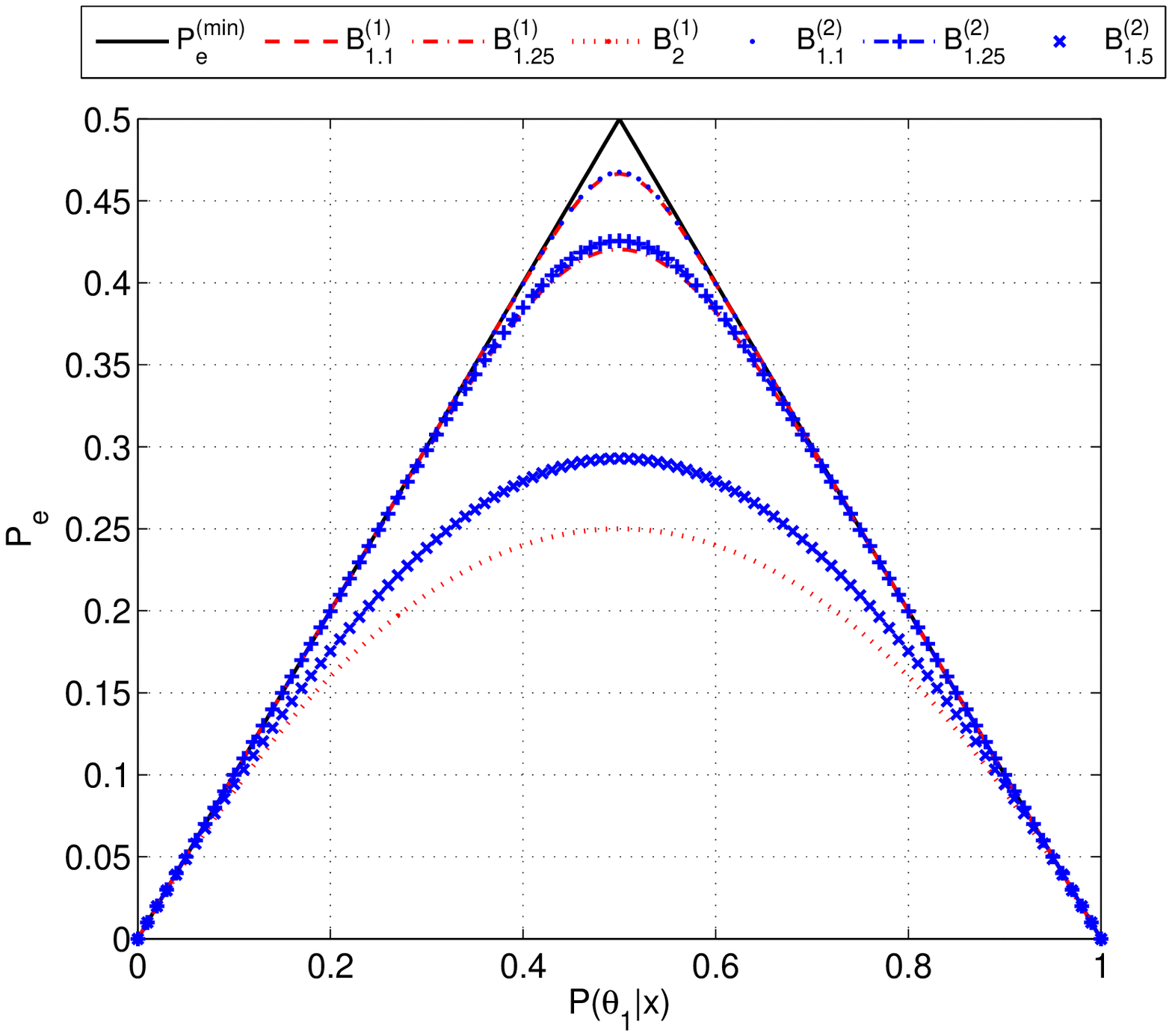,width=14cm}}
\vspace{-0.5cm}
 \caption {The  proposed lower bounds, $B_p^{(1)}$ and $B_p^{(2)}$ with $p=2,1.25,1.1$  as a function of the conditional probability
$P(\theta_1|\xvec)$ for binary
hypothesis testing.} \label{tring_new}
\end{figure}

Fig. \ref{tring_new}  depicts the   lower bounds $B_p^{(1)}$ and $B_p^{(2)}$, presented in (\ref{final_bound}) and (\ref{final_bound2}), for the binary hypothesis problem against the conditional probability $P(\theta_1|\xvec)$, for  different values of the parameter $p$ and given $\xvec$. 
It can be seen that the bounds in (\ref{final_bound}) and (\ref{final_bound2}) become  tighter as $p$ decreases and that for given $p$,  $B_p^{(2)}$ is always tighter than $B_p^{(1)}$.

\begin{figure}[htb]
\centerline{\psfig{figure=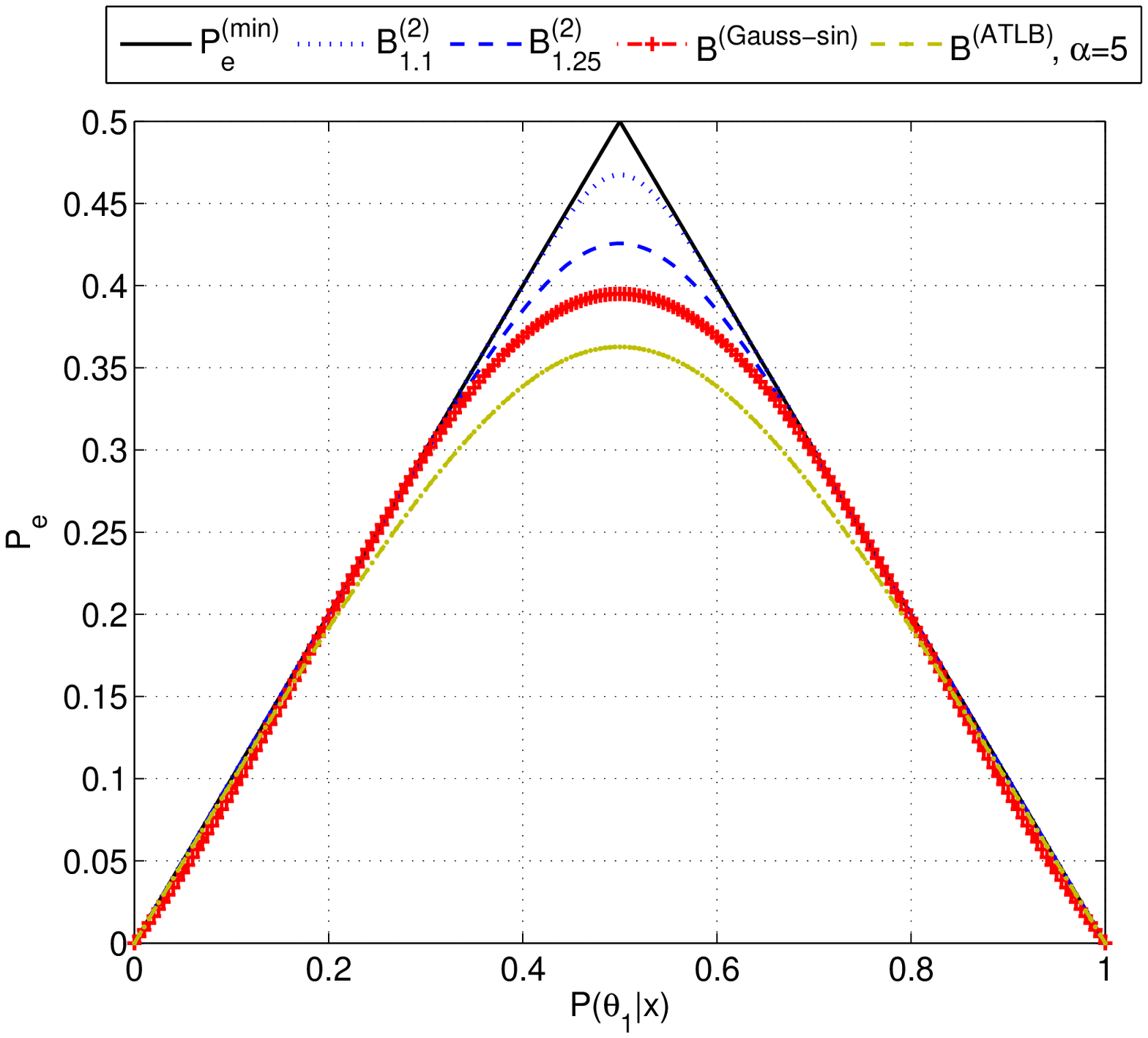,width=14cm}}
\vspace{-0.5cm}
 \caption {The  proposed lower bounds, $B_p^{(1)}$ and $B_p^{(2)}$ with $p=1.25,1.1$ compared to other existing bounds as a function of the conditional probability
$P(\theta_1|\xvec)$ for binary
hypothesis testing.} \label{tring}
\end{figure}

Fig. \ref{tring}  depicts the   lower bound $B_p^{(2)}$, presented in  (\ref{final_bound2}), for the binary hypothesis problem against the conditional probability $P(\theta_1|\xvec)$, for  different values of the parameter $p$ and given $\xvec$. The new bound is compared to the bounds $B^{(Gauss-sin)}$ and $B^{(ATLB)}$  with $\alpha=5$, presented in Appendix B.
It can be seen that $B_p^{(2)}$ becomes  tighter as $p$ decreases, and that for $p=1.1$, the new bound is tighter than  the other lower bounds almost everywhere. 

\subsection{Example: Binary hypothesis problem}
\begin{figure}[htb]
\centerline{\psfig{figure=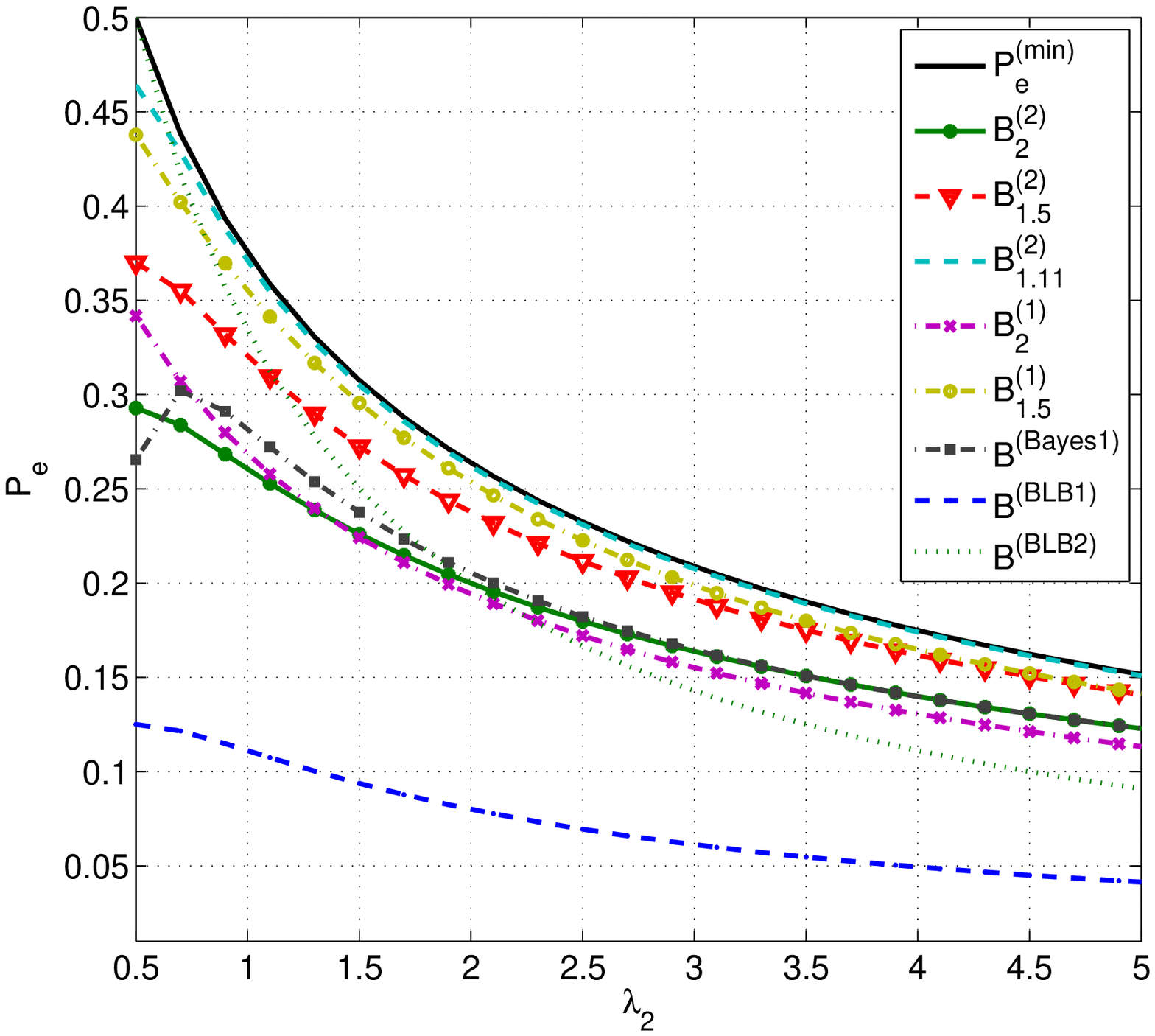,width=16cm}}
\caption {Comparison of the different lower bounds and the exact
minimum probability of error as a function of $\lambda_2$ for two equally-likely exponential distribution hypotheses.}
\label{exp}
\end{figure}

\begin{figure}[htb]
\centerline{\psfig{figure=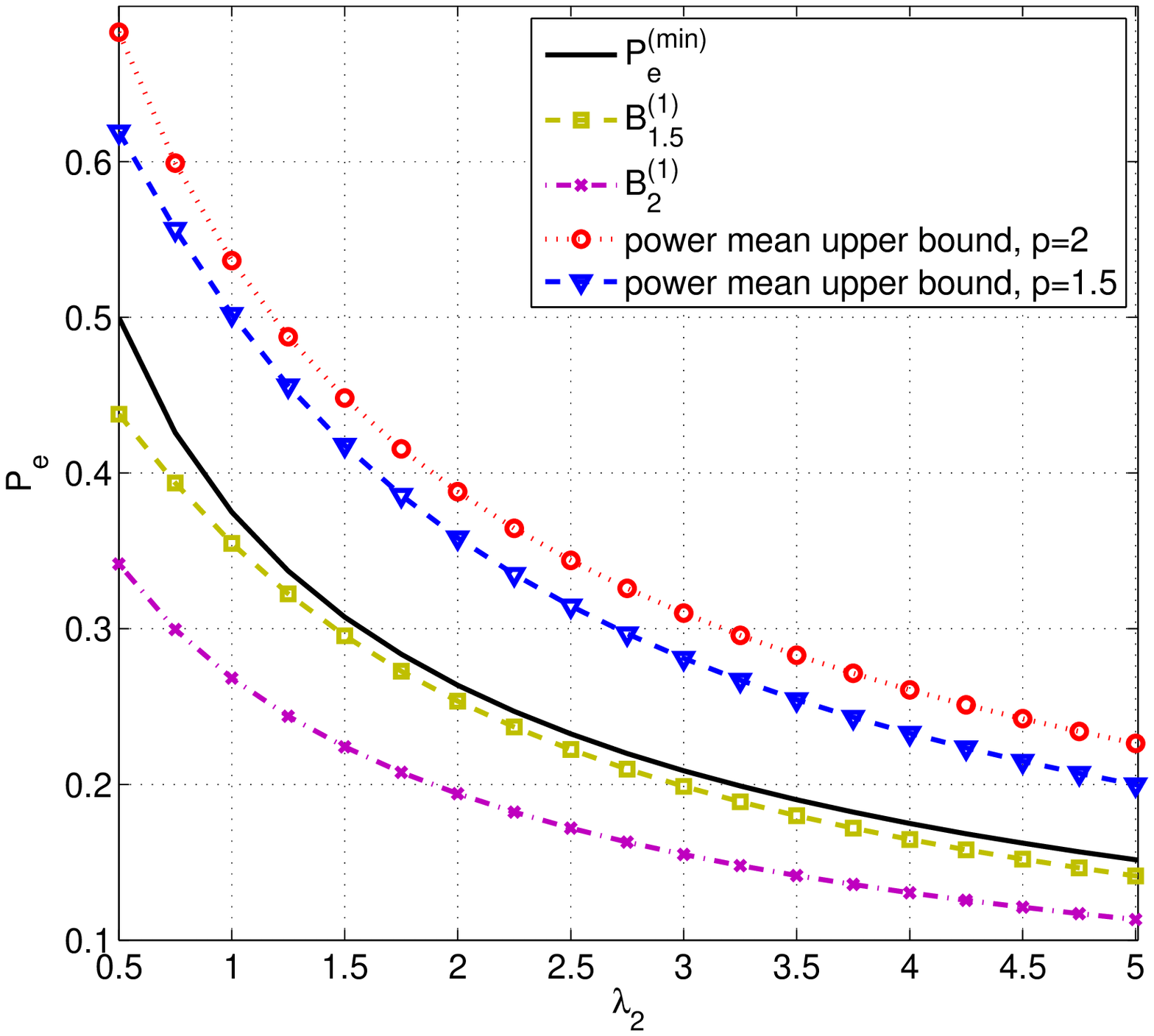,width=16cm}}
\caption {Comparison of the upper and lower bounds and the exact
minimum probability of error as a function of $\lambda_2$ for two equally-likely exponential distribution hypotheses.}
\label{exp_up}
\end{figure}

Consider the following binary hypothesis testing
problem:
\begin{equation}
{\begin{array}{rl}
&\theta_1:  f(x|\theta_1)=\lambda_1 e^{-\lambda_1 x}u(x)\\
&\theta_2:  f(x|\theta_2)=\lambda_2 e^{-\lambda_2 x}u(x)
\end{array}}
\end{equation}
where $u(\cdot)$ denotes the unit step function,
 $P(\theta_1)=P(\theta_2)=\frac{1}{2}$, and $\lambda_1=\frac{1}{2},\lambda_2>\lambda_1$.
For this problem, the  bounds in (\ref{final_bound}) with $p=2$ and $p=1.5$ are
\[{
 {B_{2}^{(1)}=\frac{1}{2}~{ _2}F_1\left( -\frac{\lambda_2}{\lambda_1-\lambda_2},1;\frac{\lambda_1-2\lambda_2}{\lambda_1-\lambda_2};-\frac{\lambda_2}{\lambda_1}\right)}}\]
\[B_{1.5}^{(1)}=\frac{1}{2}~{ _2}F_1\left( -\frac{\lambda_2}{2(\lambda_1-\lambda_2)},\frac{1}{2};1-\frac{\lambda_2}{2(\lambda_1-\lambda_2)};-\frac{\lambda_2^2}{\lambda_1^2}\right)\]
and the  bounds in (\ref{final_bound2})  are
\[B_{\frac{q}{q-1}}^{(2)}=1-\frac{1}{2}~{ _2}F_1\left( -\frac{\lambda_1}{q(\lambda_1-\lambda_2)},-\frac{1}{q};1-\frac{\lambda_1}{q(\lambda_1-\lambda_2)};-\frac{\lambda_2^q}{\lambda_1^q}\right)~~~\forall q=2,3,\ldots\]
where ${ _2}F_1$ is the hypergeometric function \cite{abramowitz_stegun}.
Several  bounds on the probability of error and the minimum probability of error obtained by the MAP detector  are presented in Fig. \ref{exp}
as a function of the distribution parameter, $\lambda_2$. The bounds depicted in this figure are: $B^{(BLB1)}$, $B^{(BLB2)}$, $B^{(Bayes1)}$ in addition to the proposed  lower bounds $B_{p}^{(1)}$ with $p=1.5,2$ and $B_{p}^{(2)}$ with $p=1.11,1.5,2$.  It can be seen that $B^{(BLB1)}$ is lower than any proposed lower bound. In addition,
for $\lambda_2\geq 0.65$ the proposed bound $B_{1.11}^{(2)}$ is tighter than all the other bounds and it is close to the minimum probability of error obtained by the MAP decision rule.
For $\lambda_2\geq 0.8$  $B_{1.5}^{(1)}$ is tighter than  the Bhattacharyya
lower bounds and   $B_{2}^{(1)}$ and $B_{2}^{(2)}$  are tighter than the $B^{(BLB1)}$ 
 everywhere and tighter than  other bounds in some specific regions. 
Fig. \ref{exp_up} presents the proposed  lower bounds $B_{p}^{(1)}$ with $p=1.5,2$  as a function of  $\lambda_2$ compared to the upper bounds on the MAP probability of error
\cite{Renyis}, given by (\ref{upper bound}). It can be seen that this class of upper bounds is proportional to the proposed tightest subclass of lower bounds in (\ref{final_bound}) with a factor of $2^{p-1}$.

\subsection{Example: Multiple hypothesis problem}
 Consider the following multiple hypothesis testing
problem:

\begin{equation}
{\begin{array}{rl}
&\theta_1:  f(x|\theta_1)=\frac{2}{3}\cos^2(x/2)e^{-|x|}\\
&\theta_2:  f(x|\theta_2)=2\sin^2(x/2)e^{-|x|}\\
&\theta_3:  f(x|\theta_3)=\frac{5}{4}\sin^2(x)e^{-|x|}
\end{array}}
\end{equation}
with $P(\theta_1)=\frac{15}{28}$, $P(\theta_2)=\frac{5}{28}$, and
$P(\theta_3)=\frac{8}{28}$. In this problem, the exact probability of error of the MAP detector is difficult to compute.
The bounds $B^{(Bayes1)}$, $B^{(Bayes2)}$, $B^{(Bayes3)}$, and $B^{(quad)}$ are not tractable. The proposed bound with $q=2$ is computable and  is equal to
\begin{eqnarray}
B_{q=2}^{(1)}=\frac{40}{14}\int_0^\infty\frac{e^{-x}}{\frac{1}{\cos^2(x/2)}+\frac{1}{\sin^2(x/2)}+\frac{1}{\sin^2(x)}}dx=\nonumber \\ =\frac{2}{35}e^{-x}\left( \cos(2x)-2\sin(2x)-5\right)|_0^\infty=0.2286 \;.\nonumber
\end{eqnarray}
 This example demonstrates  the simplicity of the proposed bound with $q = 2$,  while the other bounds are intractable.
 \section{Conclusion}
\label{diss}
 In this paper,  new classes of lower bounds on the probability of error in multiple hypothesis testing were presented.  The proposed classes depend on a parameter, $p$, which at the limit of $p\rightarrow 1^+$ approach the minimum attainable probability of  error provided by the  MAP detector.  It is shown that these classes of bounds generalize some existing bounds
for binary and multiple hypothesis testing. New variations using the proposed classes. 
It was shown via examples that the proposed bounds outperform other existing bounds in terms of tightness and simplicity of calculation.

\appendix
\section*{A. Necessary and sufficient condition for independency of (\ref{LB}) and (\ref{LB2}) on $\hat{\theta}$}
\label{Appendix A}
 In this appendix, it is shown that  the expectation  ${\rm E}\left[ \left|u(\xvec,\theta)v_k\left(\xvec,\theta\right)\right|\right], ~k=1,2$ is independent of the detector $\hat{\theta}$ {\em{iff}}%
 \begin{equation}
\label{Vdef_app}
 v_k(\xvec,\theta_i)=\frac{{\zetavec_k}(\xvec)}{P(\theta_i|\xvec)}, ~~~k=1,2,~~i=1,\ldots,M
\end{equation}
where $\zetavec_1(\cdot)$ and $\zetavec_2(\cdot)$ are  arbitrary functions of the observations $\xvec$.

Sufficient condition:  By substituting the function $ v_k(\xvec,\theta_i)=\frac{{\zetavec_k}(\xvec)}{P(\theta_i|\xvec)}$  for almost every $\xvec$ in ${\rm E}\left[ \left|u(\xvec,\theta)v_k\left(\xvec,\theta\right)\right|\right]$, one obtains
\be
\label{uv_suf}
{\rm E}\left[ \left|u(\xvec,\theta)v_k\left(\xvec,\theta\right)\right|\right]=
{\rm E}\left[ \sum _{i=1}^M  u(\xvec,\theta_i)  \zetavec_k(\xvec)\right]={\rm E}\left[ \sum _{\stackrel{i=1}{\hat{\theta}\neq\theta_i}}^M    \zetavec_k(\xvec)\right] =(M-1){\rm E}\left[\zetavec_k(\xvec) \right]
\ee
which is independent of the detector $\hat{\theta}$. Substitution of (\ref{uv_suf})
in (\ref{LB}) and (\ref{LB2}) results in bounds 
that
are independent of the detector $\hat{\theta}$.

 Necessary  condition:  Let  ${\rm E}\left[ \left|u(\xvec,\theta)v_k\left(\xvec,\theta\right)\right|\right]$ be independent of the detector $\hat{\theta}$ and  define the following sequence of two-hypothesis detectors
\be
\label{46}
\hat{\theta}_{A,j,l}(\xvec)=
\theta_j {\mathbf{1}}_{\xvec \in A}+\theta_l {\mathbf{1}}_{\xvec \in A^c}=\left\{\begin{array}{cc}\theta_j &\text{if } \xvec\in A\\
\theta_l &\text{if } \xvec\in A^c\end{array}\right.,~~~j,l=1,\ldots,M
\ee
where  $\xvec$ is a  random observation vector   with positive probability measure and $A^c$ is the complementary event of $A$.  For each detector 
\beqna
\label{random_x}
{\rm E}\left[ \left|u(\xvec,\theta)v_k\left(\xvec,\theta\right)\right|\right]=\hspace{13cm}\nonumber\\={\rm E}\left[\left.\sum_{i=1} ^M P(\theta_i|\xvec)v_k\left(\xvec,\theta_i\right) {\mathbf{1}}_{\hat{\theta}\neq\theta_i}\right|\xvec \in A\right]P(\xvec \in A)+{\rm E}\left[\left.\sum_{i=1} ^M P(\theta_i|\xvec)v_k\left(\xvec,\theta_i\right){\mathbf{1}}_{\hat{\theta}\neq\theta_i} \right|\xvec \in A^c\right]P(\xvec \in A^c)\;.
\eeqna
Using (\ref{46}) and the law of total probability, one obtains
\beqna
{\rm E}\left[ \left|u(\xvec,\theta)v_k\left(\xvec,\theta\right)\right|\right]=\hspace{-3cm}&
\nonumber\\&=&{\rm E}\left[\left.\sum_{\stackrel{i=1}{i\neq j}} ^M P(\theta_i|\xvec)v_k\left(\xvec,\theta_i\right)\right|\xvec \in A\right]P(\xvec \in A)+{\rm E}\left[\left.\sum_{\stackrel{i=1}{i\neq l}} ^M P(\theta_i|\xvec)v_k\left(\xvec,\theta_i\right) \right|\xvec \in A^c\right]P(\xvec \in A^c)
\nonumber\\
&=&{\rm E}\left[\left.\sum_{\stackrel{i=1}{i\neq j}} ^M P(\theta_i|\xvec)v_k\left(\xvec,\theta_i\right)\right|\xvec \in A\right]P(\xvec \in A)-{\rm E}\left[\left.\sum_{\stackrel{i=1}{i\neq l}} ^M P(\theta_i|\xvec)v_k\left(\xvec,\theta_i\right) \right|\xvec \in A\right]P(\xvec \in A)+\nonumber\\&+&{\rm E}\left[\left.\sum_{\stackrel{i=1}{i\neq l}} ^M P(\theta_i|\xvec)v_k\left(\xvec,\theta_i\right) \right|\xvec \in A\right]P(\xvec \in A)+{\rm E}\left[\left.\sum_{\stackrel{i=1}{i\neq l}} ^M P(\theta_i|\xvec)v_k\left(\xvec,\theta_i\right) \right|\xvec \in A^c\right]P(\xvec \in A^c)
\nonumber\\
&=&{\rm E}\left[\left.\left(\sum_{\stackrel{i=1}{i\neq j}} ^M P(\theta_i|\xvec)v_k\left(\xvec,\theta_i\right)-\sum_{\stackrel{i=1}{i\neq l}} ^M P(\theta_i|\xvec)v_k\left(\xvec,\theta_i\right)\right)\right|\xvec \in A\right]P(\xvec \in A)+{\rm E}\left[\sum_{\stackrel{i=1}{i\neq l}} ^M P(\theta_i|\xvec)v_k\left(\xvec,\theta_i\right)\right]
\nonumber\\
&=&{\rm E}\left[\left.\left(P(\theta_l|\xvec)v_k\left(\xvec,\theta_l\right)-P(\theta_j|\xvec)v_k\left(\xvec,\theta_j\right)\right)\right|\xvec \in A\right]P(\xvec \in A)+{\rm E}\left[\sum_{\stackrel{i=1}{i\neq l}} ^M P(\theta_i|\xvec)v_k\left(\xvec,\theta_i\right)\right]\;.
 \eeqna
Under the assumption that ${\rm E}\left[ \left|u(\xvec,\theta)v_k\left(\xvec,\theta\right)\right|\right]$ is independent of the detector $\hat{\theta}$,  in particular, (\ref{random_x}) is identical 
for all $j,l=1,\ldots,M$, that is this term is independent of $A$, $\theta_j$, and $\theta_l$. Thus, for given $\theta_l$, the term ${\rm E}\left[\left.\left(P(\theta_l|\xvec)v_k\left(\xvec,\theta_l\right)-P(\theta_j|\xvec)v_k\left(\xvec,\theta_j\right)\right)\right|\xvec \in A\right]P(\xvec \in A)$
 is identical for every $A$ and $\theta_j$. In particular, by setting
		 $A=\emptyset$ where $\emptyset$ is the empty set, one obtains\\ \[{\rm E}\left[\left.\left(P(\theta_l|\xvec)v_k\left(\xvec,\theta_l\right)-P(\theta_j|\xvec)v_k\left(\xvec,\theta_j\right)\right)\right|\xvec \in \emptyset\right]P(\xvec \in \emptyset)=0,~~~j=1,\ldots,M\] and therefore \[{\rm E}\left[P(\theta_l|\xvec)v_k\left(\xvec,\theta_l\right)-P(\theta_j|\xvec)v_k\left(\xvec,\theta_j\right) |\xvec \in A\right]P(\xvec \in A)=0,~~~\forall A,~j=1,\ldots,M\]
which is possible only if 
 \[
P(\theta_l|\xvec)v_k\left(\xvec,\theta_l\right)-P(\theta_j|\xvec)v_k\left(\xvec,\theta_j\right)=0,~~~\forall j=1,\ldots,M\;.
\]
Because $l$ is arbitrarily chosen, one obtains
\[
P(\theta_l|\xvec)v_k\left(\xvec,\theta_l\right)=P(\theta_j|\xvec)v_k\left(\xvec,\theta_j\right)=\zeta_k(\xvec),~~~\forall j,l=1,\ldots,M
\]
where $\zeta_k(\xvec)$ does not depend on the hypothesis.

 %%%%%%%%%%%%%%%%%%%%%%%%%%%%%%%%%%%%%%%%%%%%%%%%%%%%%%%%%%%%%%%%%%%%%%%%%%%%
\section*{B. Review of Existing Lower Bounds}
\label{exs}
In this appendix, some existing lower bounds on the minimum probability of error are presented. Part of these bounds are presented also in the review in \cite{Chen}.

{\bf{Binary hypothesis testing bounds}}\\
 Most of the binary hypothesis testing bounds are based on
 divergence measures  of the difference between two probability distributions,
 known as $f$-divergences or Ali-Silvey distances \cite{AliSilvey}.
In \cite{tutorial}, the divergence  and  two Bhattacharyya-based  lower bounds were proposed. The divergence lower bound is
\begin{equation}
\label{div1}
P_e\geq B^{(div)}=\frac{1}{8}e^{-J/2}
\end{equation}
where
$J = {\rm E}[\log L(\xvec)|\theta_1] - {\rm E}[\log L(\xvec)|\theta_2]$ and $L(\xvec)=\frac{P(\theta_1|\xvec) P(\theta_2)}{P(\theta_2|\xvec) P(\theta_1)}$ is the likelihood ratio function.
A simple Bhattacharyya-based  lower bound is
\begin{eqnarray}
\label{Bhat1}
P_e\geq B^{(BLB1)}=\frac{{\rm E}^2\left[\sqrt{P(\theta_1|\xvec)P(\theta_2|\xvec)}\right]}{8{P(\theta_1)P(\theta_2)}}\;.
\end{eqnarray}
This bound is always tighter than the divergence lower bound \cite{tutorial}.
The second Bhattacharyya-based bound on $P_e$  is
\begin{eqnarray}
\label{Bhat2}
P_e\geq B^{(BLB2)}= \frac{1}{2}-\frac{1}{2}\sqrt{1-4{\rm E}^2\left[\sqrt{P\left(\theta_1|\xvec\right) P\left(\theta_2|\xvec\right)}\right]}\;.
\end{eqnarray}
Another $f$-divergence bound is proposed in \cite{Boekee86}:
\begin{eqnarray}
\label{div}
P_e\geq B^{(f)}=\frac{1}{2}-\frac{1}{2}\sqrt{1-{\rm E}\left[(4P\left(\theta_1|\xvec\right) P\left(\theta_2|\xvec\right))^L\right]}
\end{eqnarray}
where $L\geq 1$. For $L=1$ this bound can be obtained also by applying  Jensen's inequality
on the MAP probability of error. %Note that for $L\rightarrow\infty$ this bound goes to $0$.
The harmonic lower bound was proposed in \cite{Renyis}:
\begin{equation}
\label{HMB}
P_e\geq B^{(HLB)} ={\rm E}\left[P(\theta_1|\xvec)P(\theta_2|\xvec)\right]\;.
\end{equation}

The pairwise separability measure, $J_\alpha(\theta|\xvec)={\rm E}\left[\left| P\left(\theta_1|\xvec\right)-P\left(\theta_2|\xvec\right)\right|^\alpha\right]$, is used to derive the following binary bound \cite{Jalpha}
\begin{eqnarray}
\label{Ja}
P_e\geq B^{(Ja)}=\frac{1}{2}-\frac{1}{2}J_\alpha^{\frac{1}{\alpha}}(\theta|\xvec),~~~1\geq\alpha\;.
\end{eqnarray}
The ``Gaussian-Sinusoidal'' lower bound \cite{Hashlamoun_doc} is given by
\begin{eqnarray}
\label{sin_exp}
P_e\geq B^{(Gauss-sin)}=0.395{\rm E}\left[\sin(\pi P\left(\theta_1|\xvec\right))e^{-\alpha\left(P\left(\theta_1|\xvec\right)-\frac{1}{2}\right)^2}\right]
\end{eqnarray}
where $\alpha=1.8063$. Although this bound is  tight, it is usually not tractable. An arbitrarily tight lower bound \cite{Avi96}
is given by
\begin{eqnarray}
\label{ATLB}
P_e\geq B^{(ATLB)}=\frac{1}{\alpha}{\rm E}\left[\log \frac{1+e^{-\alpha}}{e^{-\alpha P(\theta_1|\xvec)}+e^{-\alpha P(\theta_2|\xvec)}}\right]
\end{eqnarray}
for any $\alpha>0$. 
 By selecting high enough values for $\alpha$, this lower bound
can be made arbitrarily close to  $P_e^{(min)}$. However, in general this bound is difficult to evaluate.

{\bf{Multiple hypothesis testing bounds}}\\
For multiple hypothesis testing problems, the following lower bounds have been proposed.
In \cite{pattern}, Devijver derived
the following bounds using the conditional Bayesian distance:
\begin{equation}
\label{bayesDis1}
P_e\geq B^{(Bayes1)}=\frac{M-1}{M}\left(1-\sqrt{\frac{M\times{\rm E}\left[\sum\limits_{i=1}^{M}P^2\left(\theta_i|\xvec\right)\right]-1}{M-1}}\right)
\end{equation}
and
\begin{equation}
\label{bayesDis2}
P_e\geq B^{(Bayes2)}=1-\sqrt{{\rm E}\left[\sum\limits_{i=1}^{M}P^2\left(\theta_i|\xvec\right)\right]}
\end{equation}
where ${\rm E}\left[\sum\limits_{i=1}^{M}P^2\left(\theta_i|\xvec\right)\right]$ is the conditional Bayesian distance. The bound in (\ref{bayesDis1}) with $M=2$ is identical to (\ref{Ja}) with $\alpha=2$.
In \cite{pattern}, it is analytically shown  that for the binary case
the Bayesian distance lower bound in (\ref{bayesDis1}) is always tighter than the Bhattacharyya bound in (\ref{Bhat2}).
Using Jensen's inequality, the following bound  is tighter than the bound in (\ref{bayesDis2})
\cite{Renyis}, \cite{pattern}
\begin{equation}
\label{bayesDis3}
P_e\geq B^{(Bayes3)}= 1-{\rm E}\left[\sqrt{\sum_{i=1}^{M}P^2(\theta_i|\xvec)}\right]\;.
\end{equation}
The bound
\begin{equation}
\label{har_2}
P_e\geq B^{(quad)}=\frac{1}{2}-\frac{1}{2}{\rm E}\left[\sum\limits_{i=1}^{M}P^2\left(\theta_i|\xvec\right)\right]
\end{equation}
was  proposed in  \cite{TOUdoc} and \cite{Vaj68} in the context of Vajda's
quadratic entropy and
the quadratic mutual information, respectively. Note that
the bound $B^{(quad)} $ can
be interpreted as an $M$-ary extension to the harmonic mean
bound, presented in (\ref{HMB}). In \cite{pattern}, it is claimed that
$
B^{(quad)} \leq B^{(Bayes2)} \leq B^{(Bayes1)} \;.
$ 
The affinity  measure of  information relevant to the discrimination
among the $M$ hypothesis is defined
as lower bound on $P_e$
\cite{Matusita}
\begin{equation}
\label{MatusitaLB}
P_e\geq B^{(MLB)}=\frac{M-1}{M^{M-1}}\left({\rm E}\left[\prod_{i=1}^M P^{\frac{1}{M}}(\theta_i|\xvec) \right]\right)^M\;.
\end{equation}
%The average quadratic entropy, $h(\theta|\xvec)\define {\rm E}\left[\sum_{i=1}^M P\left(\theta_i|\xvec\right)\left(1-P\left(\theta_i|\xvec\right)\right)\right]$ is used to derive the following bound \cite{}
%\begin{equation}
%P_e\geq B^{(quad2)}=\frac{h(\theta|\xvec)}{1+\sqrt{1-2h(\theta|\xvec)}}\;.
%\end{equation}

The ``general mean distance" between the $M$ hypotheses is 
$G_{\alpha,\beta}={\rm E}\left[ \left(\sum_{i=1}^M P^{\beta}\left(\theta_i|\xvec\right)\right)^\alpha \right]$ \cite{boekee79}, \cite{Basseville89}.
Many lower bounds on $P_e$ and upper bounds on $P_e^{(min)}$ can be obtained from this distance \cite{Basseville89}. For example, the binary bound in (\ref{upper bound}) and the following classes of bounds: 
\beqna
\label{Gab2}
P_e\geq B^{(GMD1)}= 1-G_{\alpha,\beta}^{\frac{1}{\alpha \beta}},~~~0<\alpha,~~~1<\beta,~~~\frac{1}{\alpha}\leq \beta 
\eeqna
\beqna
\label{Gab4}
P_e \geq B^{(GMD2)}=1-G_{\alpha,\beta},~~~0<\alpha,~~~1<\beta\leq\frac{1}{\alpha}
\eeqna
It can be seen that  by substituting $\beta=2,\alpha=1$ in (\ref{Gab2})   we obtained the lower bounds in (\ref{bayesDis2}).
By substituting $0<\alpha<1$ and $\frac{1}{\alpha}=\beta$ in (\ref{Gab2}) or  in   (\ref{Gab4}), one obtains the  lower bound
\beqna
\label{Gab_frac2}
P_e \geq 1-{\rm E}\left[ \left(\sum_{i=1}^M P^{\beta}\left(\theta_i|\xvec\right)\right)^{\frac{1}{\beta}} \right] \;.
\eeqna

\section*{Acknowledgment}
This research was supported by THE ISRAEL SCIENCE 
FOUNDATION (grant No. 1311/08).

\bibliographystyle{IEEEtran}
\bibliography{errorbound4}
\end{document}